\documentclass[twocolumn, letter, superscriptaddress,
  showpacs]{revtex4-1} 

\usepackage{amsmath, array} 
\usepackage{graphicx}
\usepackage{hyperref} 
\usepackage{color}

\newcommand{\EqLabel}[1]{\label{#1}}

\begin{document}

\title{Non-Zhang-Rice singlet character of the first ionization state of T-CuO}

\author{Clemens P. J. Adolphs} \affiliation{\!Department \!of \!Physics and
  Astronomy, \!University of\!  British Columbia, \!Vancouver, British
  \!Columbia,\! Canada,\! V6T \!1Z1} 
\author{Simon Moser} \affiliation{Institute of Condensed Matter
  Physics, Ecole Polytechnique Federale de Lausanne, CH-1015 Lausanne,
  Switzerland} 
\affiliation{Advanced Light Source, Berkeley, California 94720, USA} 
\author{George A. Sawatzky}
\affiliation{\!Department \!of \!Physics and Astronomy, \!University
  of\!  British Columbia, \!Vancouver, British \!Columbia,\! Canada,\!
  V6T \!1Z1} \affiliation{\!Quantum Matter \!Institute, \!University
  of British Columbia, \!Vancouver, British \!Columbia, \!Canada,
  \!V6T \!1Z4}
\author{Mona Berciu}
\affiliation{\!Department \!of \!Physics and Astronomy, \!University
  of\!  British Columbia, \!Vancouver, British \!Columbia,\! Canada,\!
  V6T \!1Z1} \affiliation{\!Quantum Matter \!Institute, \!University
  of British Columbia, \!Vancouver, British \!Columbia, \!Canada,
  \!V6T \!1Z4}

\begin{abstract}
We argue that tetragonal CuO (T-CuO) has the potential to finally
settle long-standing modelling issues for cuprate physics. We compare
the one-hole quasiparticle ({\it qp}) dispersion of T-CuO to that of
cuprates, in the framework of the strongly-correlated
($U_{dd}\rightarrow \infty$) limit of the three-band Emery
model. Unlike in CuO$_2$, magnetic frustration in T-CuO breaks the
C$_4$ rotational symmetry and leads to strong deviations from the
Zhang-Rice singlet picture in parts of the reciprocal space. Our
results are consistent with angle-resolved photoemission spectroscopy
data but in sharp contradiction to those of a one-band model
previously suggested for them. These differences identify T-CuO as
an ideal material to test a variety of scenarios proposed for
explaining cuprate phenomenology.
\end{abstract}
\date{\today}

 \maketitle

{\em Introduction}: Understanding the high-temperature
superconductivity in cuprates \cite{Bend} is one of the biggest
challenges in condensed matter physics. These layered materials
contain two-dimensional (2D) CuO$_2$ layers which exhibit
antiferromagnetic (AFM) order when undoped, and host superconductivity
upon doping. Consequently, it is widely believed that understanding
the behaviour of a doped CuO$_2$ layer is the key to understand the
unusual properties of these materials.

The first step is to understand the nature of the quasiparticle ({\em
  qp}) that forms when a hole is doped into a CuO$_2$ layer. Despite
many efforts, this issue is not yet settled.

The Cu $3d_{x^2-y^2}$ and ligand O $2p$ are the most relevant
orbitals, and their appropriate model is the three-band Emery
Hamiltonian \cite{Emery}. Zhang and Rice argued that its quasiparticle
is a Zhang-Rice singlet (ZRS) hopping on the Cu sublattice, well
described by the (relatively) simpler one-band $t$-$J$ or Hubbard
Hamiltonians \cite{Zhang,George,amol}. Significant effort focusing on
these one-band models followed.  In the absence of exact solutions or
accurate approximations, progress came from numerical studies of
finite-size clusters and from Cluster Dynamical Mean-Field Theory
\cite{Lichtenstein00}. These showed that the {\em qp} dispersion is
strongly influenced by the quantum fluctuations of the AFM background
\cite{Trugman}, and that longer-range hopping is necessary for
quantitative agreement with experimental measurements
\cite{Leung95,Leung97,Wells,Damascelli}.  The longer-range hoppings
required to achieve this agreement agree with those calculated
theoretically \cite{ok, eskes}. This was taken as proof that these
extended one-band models are correct, and the
focus shifted to studying them at finite doping
\cite{comm}. While much work was done in the past two decades,
the lack even of consensus that they support robust,
high-temperature superconductivity raises  doubts about how
appropriate they are to describe the hole-doped cuprates \cite{comment}.

There are two reasons why one-band models might fail to capture the
desired physics at finite doping: (i) they may describe the {\em qp}
correctly yet fail to appropriately model the effective interactions
between {\em qps}, responsible for pairing.  This was shown to occur
when degrees of freedom from different sublattices are mapped onto an
effective one-band model \cite{Mirko}. Because in cuprates the doped
holes reside on oxygen whereas the magnons reside on Cu \cite{Bayo}, a
one-band model may similarly fail to mimic their full interaction;
(ii) they may predict the correct {\em qp} dispersion for the wrong
reasons. Support for the latter view comes from our recent work on the
$U_{dd}\rightarrow \infty$ limit of the Emery model; the resulting
Hamiltonian has spins at the Cu sites and doped holes on the O
sublattice \cite{Bayo}. In stark contrast to one-band models where
spin fluctuations are key to obtaining the correct {\em qp}
dispersion, here this is found even in their absence \cite{Hadi}.
This qualitative difference shows that although these {\em qps} have
similar dispersion, it is controlled by different physics
\cite{Hadinew}.

To fully decide whether these one- and three-band models are
equivalent, one must compare them for a material like CuO$_2$, so that
it is described by similar Hamiltonians, however one where they give
different predictions. In this Letter we show that tetragonal CuO
(T-CuO) is precisely this material whose investigation can finally
resolve these fundamental modeling issues.

\begin{figure}[t]
  \centering \includegraphics[width=\columnwidth]{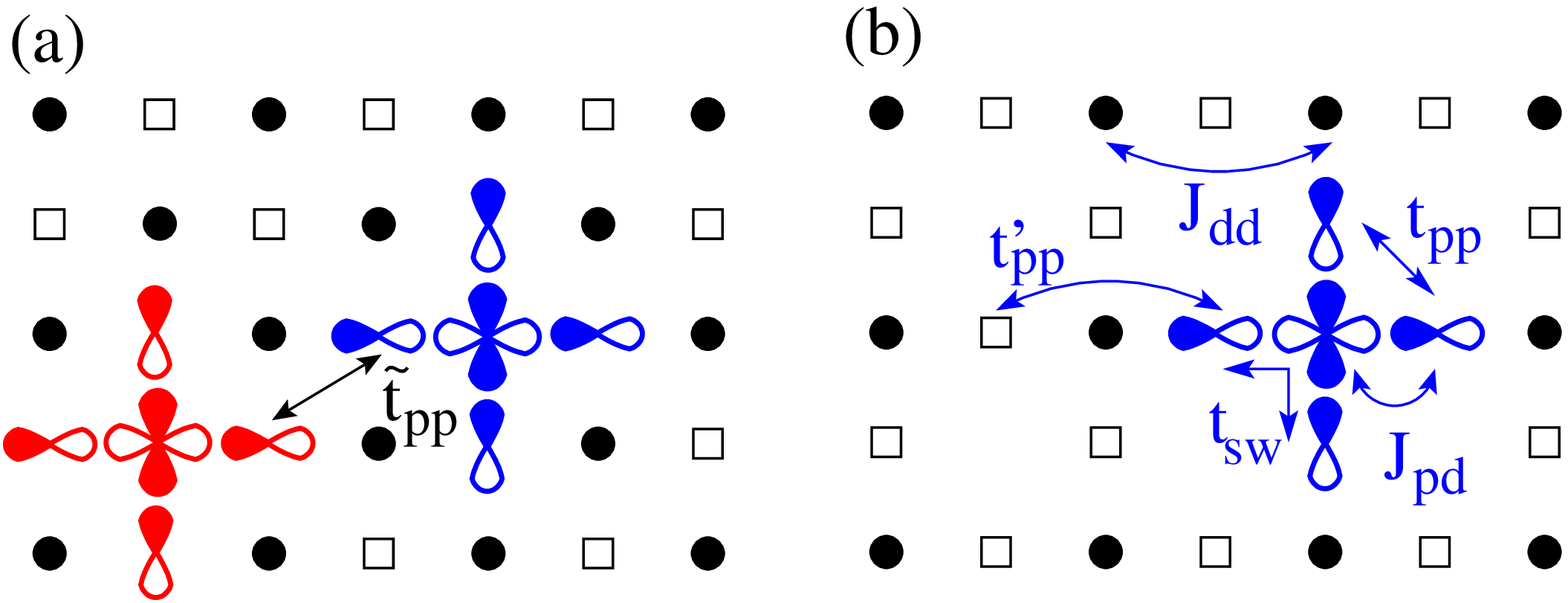}
\includegraphics[width=\columnwidth]{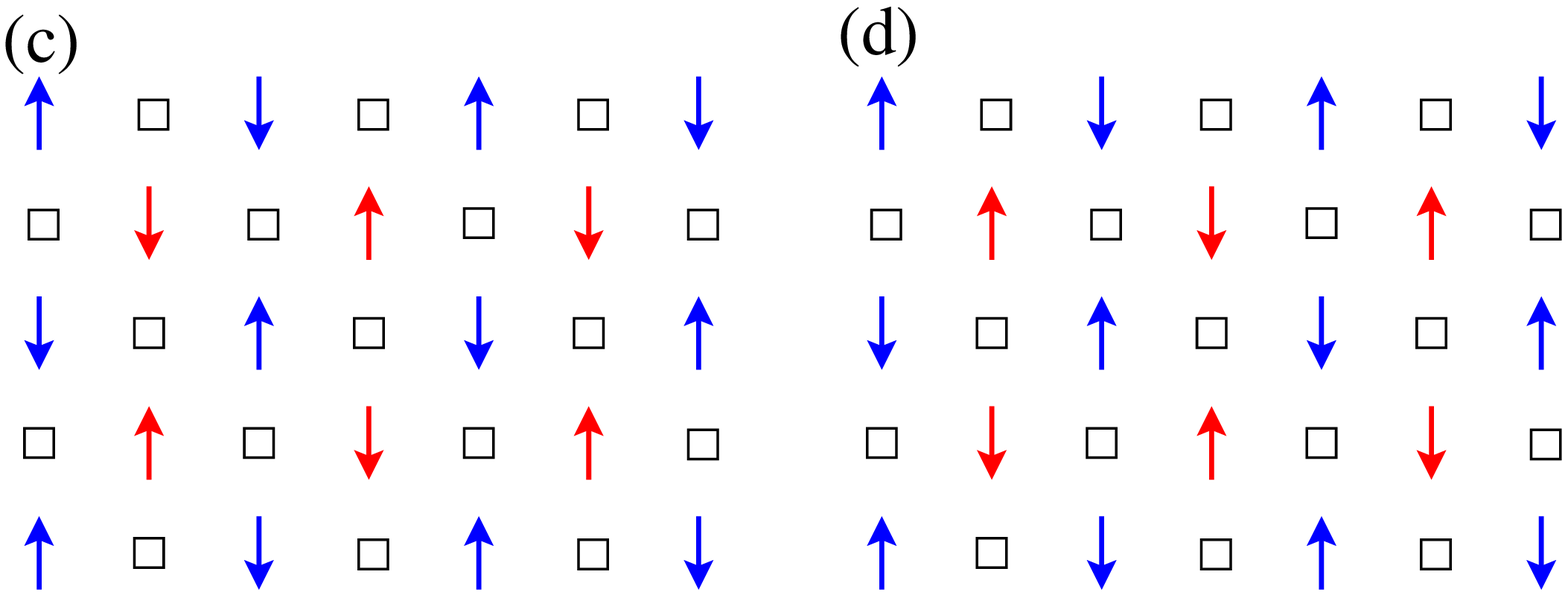}
  \caption{(color online) Structure of a layer of (a) T-CuO, and (b)
    CuO$_2$. Full circles are Cu, empty squares are O. The Cu
    $3d_{x^2-y^2}$ orbitals are drawn at a few sites, with white/dark lobes
    showing our choice for positive/negative signs. The corresponding ligand O $2p$
    orbitals are also indicated on neighboring O sites. The T-CuO
    layer can be thought of as two intercalated CuO$_2$ layers sharing common O.
    The coppers of the two sublattices hybridize with different O $2p$
    orbitals. Panels (c) and (d) show the two degenerate ground-states
    of the undoped T-CuO layer. Different colors are used for the Cu
    spins on the two sublattices for better visibility.
  \label{fig1}}
\end{figure}

Thin films of T-CuO were recently grown epitaxially on SrTiO$_3$
\cite{CuO}. They consist of stacks of weakly-interacting CuO layers,
whose structure has two intercalated CuO$_2$ lattices (sharing the
same O), see Fig. \ref{fig1}(a).  Fig. \ref{fig1}(b) shows a CuO$_2$
layer. Because Cu $3d_{x^2-y^2}$ orbitals only hybridize with their
ligand O $2p$ orbitals, shown in the same color in Fig. \ref{fig1},
the two CuO$_2$ sublattices would be effectively decoupled if $pp$
hopping between the two sets of O $2p$ orbitals was absent
\cite{commentn}. In this case, a hole doped into one sublattice would
evolve just like in a CuO$_2$ layer, and the same (but
doubly-degenerate) {\em qp} dispersion would be predicted by both one-
and three-band models, as discussed.

However, the CuO$_2$ sublattices are coupled by $pp$ hopping, which
lifts this degeneracy. The resulting {\em qp} dispersion was measured
by angle resolved photoemission spectroscopy (ARPES) \cite{grioni}. It
seems to be quite similar to that of CuO$_2$ and was argued to be well
described by the $t$-$t'$-$t''$-$J$ model \cite{grioni}.  As we show
next, this is opposite to what we find for the $U_{dd}\rightarrow
\infty$ limit of the three-band model. We predict  qualitatively
different dispersions for T-CuO and CuO$_2$, however their differences are
hidden in magnetically twinned samples.  We present our results next
and then explain why they cannot be reproduced by one-band models.

{\em Model}: We study the $U_{dd}\rightarrow \infty$ limit of the
Emery model, with spins at the Cu sites and a single doped
  hole on the O sublattice. This limit is justified because
$U_{dd}$ is the largest energy scale \cite{Ogata08}.
The corresponding Hamiltonian,
see Fig. \ref{fig1}(b), is \cite{Bayo}:
\begin{equation}
\EqLabel{1} \hat{H} = \hat{T}_{pp} + \hat{T}_{swap} +
\hat{H}_{J_{pd}}+\hat{H}_{J_{dd}}.
\end{equation}
$\hat{T}_{pp}=\sum_{i\ne j\in {\rm O}, \sigma}
t_{i-j}p^{\dagger}_{i,{\sigma}}p^{}_{j,\sigma}$ describes next-nearest
(nn) $t_{pp}$, and 2nd nn $t'_{pp}$ hopping of the hole between ligand
O $2p$ orbitals; the latter is restricted to oxygens bridged by a
Cu. For technical details see the
Supplemental Material \cite{supp} and Ref.  \cite{comm3}. $T_\mathrm{swap}$  describes Cu-mediated  hopping accompanied by a
spin-swap. Specifically, the hole at a Cu site adjacent to the
doped hole hops to another neighbor O, followed by the
doped hole falling into the vacated Cu orbital. Because the original
doped hole replaces the Cu hole, their spins are swapped.  Thus
$T_\mathrm{swap} = -t_{sw}\sum_{i \in {\rm Cu}, {\bf u \ne u'},
  \sigma, \sigma'} s_{\bf u-u'}p^{\dagger}_{i+ \bf
  u,{\sigma}}p^{}_{i+\bf u',\sigma'}|i_{\sigma'}\rangle\langle
i_\sigma|$, where ${\bf u},{\bf u'} = (\pm 0.5,0), (0, \pm 0.5)$ are
the distances between Cu and its nn O sites. It shows the
change of the Cu spin located at ${\bf R}_i$ from $\sigma$ to
$\sigma'$ as the doped hole changes its spin from $\sigma'$ to
$\sigma$ while moving to another O. The sign $s_{\boldsymbol\eta}=\pm
1$ comes from overlaps of the orbitals involved \cite{supp}.  
$\hat{H}_{J_{pd}}=J_{pd} \sum_{i, {\bf u}} {\bf S}_{i}\cdot{\bf
  s}_{i+{\bf u}}$ is generated when the Cu hole hops onto
the O hosting the doped hole, followed by one of the two holes
returning to the Cu. This gives rise to AFM exchange
between the spins ${\bf s}_{i+{\bf u}}$ of the doped hole and ${\bf S}_i$ of its neighbor Cu. Finally,
$\hat{H}_{J_{dd}}=J_{dd}\sum_{\langle i, j \rangle'} {\bf
  S}_{i}\cdot{\bf S}_{j}$ is the AFM coupling between nn Cu
spins, except on the bond blocked by the doped hole. Its energy scale
$J_{dd} \sim 150$ meV is taken as the unit of energy, in terms of
which $t_{pp}=4.1$, $t'_{pp}=0.6t_{pp}$, $t_{sw}=3.0$ and $J_{pd}=2.8$
\cite{Ogata08}. The  Hubbard repulsion $U_{pp}$ is
not included in Eq. (\ref{1}) because we consider only the case of a single doped hole.

In CuO$_2$ the ligand orbitals are the important ones, but it is
straightforward to also include the in-plane non-ligand
orbitals. These do not hybridize with Cu $3d_{x^2-y^2}$ so their
addition does not affect $\hat{T}_{swap}$, $\hat{H}_{J_{pd}}$ or
$\hat{H}_{J_{dd}}$, which arise from such hybridization.  Only
$\hat{T}_{pp}$ must be supplemented accordingly. By symmetry, nn
hopping between two non-ligand orbitals is the same $t_{pp}$ as for
ligand orbitals, with signs dictated by the lobes' overlap. Hopping
between ligand and non-ligand orbitals, denoted $\hat{T}_{mix}$ and
shown by the arrow in Fig. \ref{fig1}(a), has magnitude
$\tilde{t}_{pp}/t_{pp} =
(t_{pp,\sigma}-t_{pp,\pi})/(t_{pp,\sigma}+t_{pp,\pi})=0.6$ because
$t_{pp,\sigma} = 4 t_{pp,\pi}$ \cite{Harrison}.  For CuO$_2$,
inclusion of the non-ligand orbitals has a minor effect on the {\it
  qp} dispersion \cite{Hadi}.

The Hamiltonian for T-CuO is a straightforward generalization of
Eq. (\ref{1}). Hole hopping is described by the same $ \hat{T}_{pp}+
\hat{T}_{mix}$. Because of the two Cu sublattices, however, there are
two sets of terms $\hat{T}_{swap}$, $\hat{H}_{J_{pd}}$ and
$\hat{H}_{J_{dd}}$ which couple Cu spins on each sublattice to each
other and to the doped hole, when it occupies a $2p$ orbital with
ligand character for that sublattice. We use the same parameters for
T-CuO like for CuO$_2$ (the results remain qualitatively similar if
the parameters are varied within reasonable ranges) and focus on the
effect of $\hat{T}_{mix}$, which moves the hole between the two sets
of $2p$ orbitals and changes to which Cu sublattice it is coupled
\cite{commentn}.

\begin{figure*}
    {\includegraphics[width=1.4\columnwidth]{fig2a.eps}} \hfill \centering
    \raisebox{0.075\height}{\includegraphics[width=0.5\columnwidth]{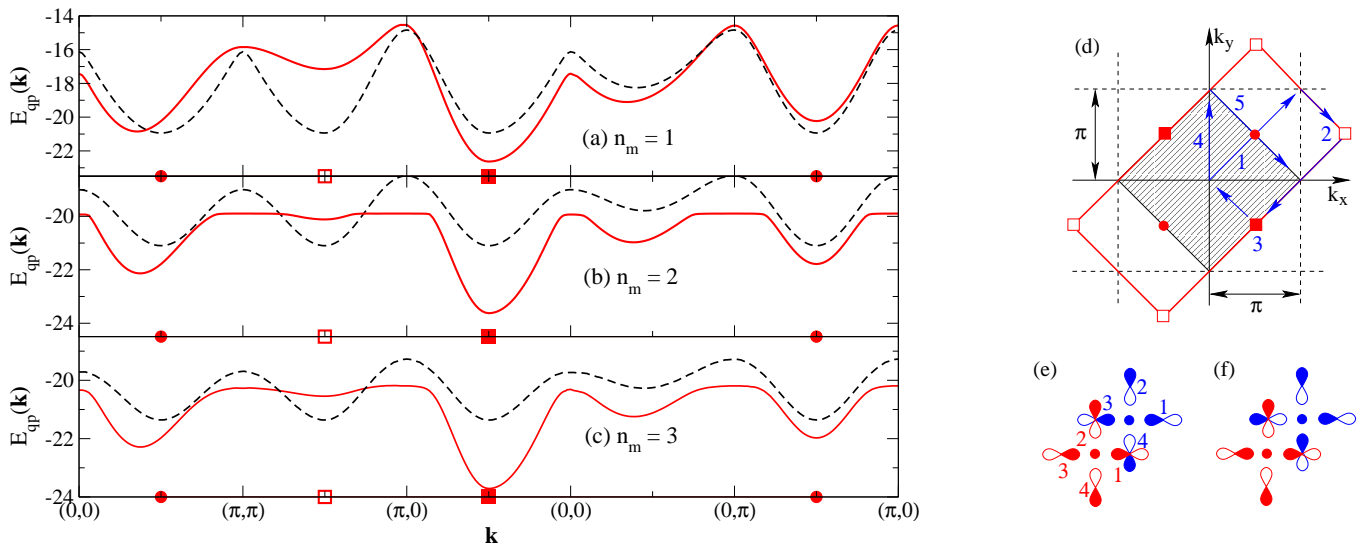}}
  \caption{(color online) $Qp$ dispersion in units of
    $J_{dd}$ for (a) $n_m=1$, (b) $n_m=2$, and (c) $n_m=3$ with full/dashed lines
    for T-CuO/CuO$_2$. The Brillouin zone for the magnetic order of
    Fig. \ref{fig1}(c) is shown in red in (b). The shaded area is the
    smaller BZ for CuO$_2$. The points marked by circles and
    empty/full squares
    are equivalent in CuO$_2$ but not in T-CuO. (d) Hopping between
    two adjacent ZRSs, and (e) between a ZRS (red) and one with $x-y$
    symmetry (blue). See text for more details.
  \label{fig2}}
\end{figure*}

{\em Variational approximation}: We extract the {\it qp} dispersion
$E_{qp}({\bf k})$ from the one-hole propagator computed variationally
in a restricted Hilbert space that allows up to $n_m$ magnons to be
created by the doped hole through $\hat{T}_{swap}$ and
$\hat{H}_{J_{pd}}$ processes, assuming that it was injected in a
N\'eel-like background \cite{supp,Hadi}. Of course, in reality there
are spin fluctuations in the AFM background, but because their energy
scale $J_{dd}$ is small, they are slow and have little effect
on the {\it qp}: the hole creates and moves its magnon
cloud on a timescale faster than that controlling the spin
fluctuations, so the latter can be ignored \cite{Hadi,Hadinew}.  If
the T-CuO energy scales are similar, and given the weak coupling
between the two Cu sublattices, this approximation should remain
valid.

In undoped T-CuO each Cu sublattice has AFM order due to its
$\hat{H}_{J_{dd}}$ term. Any weak coupling
$\tilde{J}_{dd}$ between the two Cu sublattices is therefore fully frustrated: any
spin  interacts with equal numbers of up and down
spins from the other sublattice. Nevertheless, order by disorder
selects one of the two degenerate states depicted in
Fig. \ref{fig1}(c), (d) as the ground-state of the undoped system
\cite{AFM}. Because they have FM chains running along the $x=\pm y$
diagonals, they are related by a $C_4$ rotation so it
suffices to study one case. Thus for T-CuO in either of
these states, the quasiparticle dispersion $E_{qp}({\bf
  k})$ is not invariant to $C_4$ rotations, only to $C_2$ ones.

{\em Results}: Figures
\ref{fig2}(a)-(c) show $E_{qp}({\bf k})$ from the variational method with
$n_m=1,2,3$, respectively, for the magnetic order of
Fig. \ref{fig1}(c). The Brillouin zone (BZ) is displayed in
Fig. \ref{fig2}(d). Full/dashed lines are for T-CuO/CuO$_2$.

In CuO$_2$, at the points marked by circles and squares there are
identical, nearly isotropic minima \cite{Wells,Damascelli}. With
increasing $n_m$, the bandwidth narrows and the dispersion flattens
below the polaron+one magnon continuum (both are
standard polaronic effects \cite{Hadi}) but the
shape is unchanged. The results are nearly converged at $n_m=3$
for CuO$_2$, with a bandwidth of $\sim 2J_{dd}$ in agreement with 
exact diagonalization results and experimental data \cite{Hadi}).

In T-CuO, we verified that for $\hat{T}_{mix}=0$ the same (but now
doubly-degenerate) dispersion is obtained. When $\hat{T}_{mix}$ is
turned on, this degeneracy is lifted. Only the low-energy eigenenergy
is shown in Fig. \ref{fig2}. Again, results are essentially
  converged for $n_m=3$. As expected, the dispersion loses its
invariance to $C_4$ rotations because the $qp$ now moves in a
magnetic background that lacks this symmetry.  In the
$k_x=-k_y$ quadrants $E_{qp}({\bf k})$ again displays deep, isotropic minima around $\pm
({\pi\over 2}, - {\pi\over 2})$ (full squares) and is thus 
similar to  CuO$_2$. The difference, however, is significant in
the $k_x=k_y$ quadrants near the $\pm ({\pi\over 2}, {\pi\over 2})$
points (circles). Not only are energies here higher than at the $\pm
({\pi\over 2}, - {\pi\over 2})$, but these minima are shifted
toward the $\Gamma$ point. Note that the BZ corners (empty
squares) still mark local minima, but lying at high energies
just below the polaron+one magnon continuum.

We now prove that this unusual dispersion for T-CuO involves physics
beyond the Zhang-Rice singlet. As such, it cannot be described by
one-band models obtained through a projection onto these states.

We start by estimating the effect of $\hat{T}_{mix}$ on
the CuO$_2$ degenerate eigenstates that appear in its absence,
whose energy $E_0({\bf k})$ is shown by dashed lines in
Fig. \ref{fig2}. Especially near $(\pm {\pi\over 2}, \pm {\pi\over
  2})$, the CuO$_2$ {\em qp} indeed has a large overlap
with a ZRS Bloch state \cite{Hadinew}, and the hole occupies the   ${x^2-y^2}$ linear
combination of O $2p$ ligand 
orbitals sketched for two nn sites in Fig. \ref{fig2}(e). For T-CuO, these 
degenerate states combine into one Bloch state $|d, {\bf k}\rangle$  with
momentum ${\bf k}$ in its bigger BZ. If
we  use $|d, {\bf k}\rangle$ as an
approximation for the low-energy eigenstate, then the T-CuO dispersion
becomes $E_{qp}({\bf k})\approx E_0({\bf k}) + \delta E({\bf k})$,
where $\delta E({\bf k})=\langle d, {\bf k}| \hat{T}_{mix}|
d, {\bf k}\rangle$ is:
\begin{equation}
\EqLabel{2} \delta E({\bf k}) = - \tilde{t}_{pp} \cos {k_x +
  k_y\over 2} \left[ 1- \cos(k_x-k_y)\right].
\nonumber
\end{equation}
The cosines are a geometric factor from 
 the  Bloch states' phase differences between neighboring Cu sites \cite{supp}.

Because $\delta E(k_x=-k_y)= -2\tilde{t}_{pp}\sin^2 k_x$ and $ \delta
E(k_y=k_x\mp\pi)=-2\tilde{t}_{pp}\sin |k_x|$, minima at $\pm
({\pi\over2}, -{\pi\over2})$ (full squares) move to lower energies
while minima at the BZ corners (empty squares) move up. This agrees
with the results of Fig. \ref{fig2}.

However, because $\delta E(k_x=k_y)=0$, the dispersion near
$\pm({\pi\over 2}, {\pi\over 2})$ (circles) should remain unchanged
instead of these minima moving toward the $\Gamma$ point. Moreover, we
find that the overlap between the T-CuO {\em qp} and $|d, {\bf k}\rangle$ {\em vanishes} at ${\bf k}=
\pm({\pi\over 2}, {\pi\over 2})$. These facts clearly prove that the
changes near the $\pm({\pi\over 2}, {\pi\over 2})$ points cannot be
due to Zhang-Rice singlet physics.

Indeed, $\hat{T}_{mix}$ hopping between $x^2-y^2$ linear combinations
centred at nn Cu sites is suppressed, see Fig. \ref{fig2}(e): {\em
  eg.}, a hole at site 1 of the lower Cu (red) hops into $p^\dagger_1
+ p^\dagger_3$ of the upper Cu (blue), which is orthogonal to its
$x^2-y^2$ linear combination. Instead, hopping between adjacent
$x^2-y^2$ and $x-y$ combinations is enhanced, see Fig. \ref{fig2}(f).
The shift of the $\pm({\pi\over 2}, {\pi\over 2})$ minima toward
$\Gamma$ is due to a large mixing of the singlet with $x-y$ symmetry
into the quasiparticle eigenstate, which thus loses its ZRS nature
(for more details see the Supplemental Material \cite{supp}). 
  Note that experiments like Refs. \cite{ZRSexp},
  which are sensitive only to the local singlet character, cannot
  distinguish a ZRS singlet from one with such mixed symmetry.

We checked that adding terms like $\tilde{J}_{dd}$ and
$\tilde{J}_{pd}$ \cite{commentn} has no qualitative effects on the
dispersion. This is expected because their matrix elements are small
and/or featureless near $(\pm {\pi \over 2}, \pm {\pi \over 2})$. We
are therefore confident that our prediction is robust.

 ARPES finds the T-CuO {\em qp} dispersion to obey C$_4$ symmetry and
 to have a large BZ, corresponding to a unit cell containing one Cu
 and one O atom \cite{grioni}. Both features are very surprising for
 the long-range magnetic orders of Figs. \ref{fig1}(c), (d), 
 which break the C$_4$ symmetry. Moreover, any AFM-type order has at
 least two magnetically non-equivalent Cu atoms so its BZ is
 like in Fig. \ref{fig2}(d) or smaller, never larger. Our results become
 consistent with ARPES if we assume the presence of domains
 in both ground-states, so that their average is measured
 experimentally. Indeed, as shown in the Supplemental Material
 \cite{supp}, averaging $E_{qp}({\bf k})$  of Fig. \ref{fig2}(d) with
 its counterpart rotated by 90$^o$ leads to an apparent doubling of
 the BZ and a new pattern of minima with two different
 energies, in agreement with those found experimentally.

We predict that a dispersion like in Fig. \ref{fig2} appears in
the ARPES of ``magnetically untwinned'' T-CuO films in the insulating
limit.  This is very different and thus easily distinguishable
from the one-band model prediction \cite{grioni}. The observation of
this pattern, with shallower displaced minima in two  quadrants,
will provide a clear proof of low-energy physics beyond the ZRS, and
of the superiority of three-band models to model such materials. If
T-CuO films can be doped, this new pattern of
minima will open extraordinary opportunities to test many ideas
relating the shape of the Fermi surface, location of ``hot spots'' and
possibility of nesting, to much of the cuprate phenomenology,
including the symmetry of the  superconducting gap,
formation of stripes, appearance and relevance of various other
ordered phases, etc. 

We note that ARPES measurements on untwinned pnictides have been successfully
performed (see, e.g., \cite{untwinned}). It is therefore reasonable to expect
that similar measurements for T-CuO are feasible.  An important lesson from
this study is that low-energy physics of non-ZRS nature can arise in T-CuO and
similar materials in suitable circumstances/symmetries. The presence of
disorder, of other nearby quasiparticles, of stripes, charge-density wave or
other ordered phases may have a similar effect in CuO$_2$ layers.

{\em Acknowledgements}: We thank G. Koster and M. Grioni for useful
discussions. Work was supported by NSERC, QMI, CIFAR, SNF, and a UBC
4Y Fellowship (CA).


\begin{thebibliography}{99}
\bibitem{comment} It is much more likely that the one-band models (with
properly adjusted longer-range hopping) describe correctly the physics
of the electron-doped side, because in this case the full O bands are
inert spectators.
\bibitem{commentn} For completeness, we note that in
Ref. \cite{grioni} a small nn exchange $\tilde{J}_{dd} \sim 0.04$
 was also included between Cu spins on the different sublattices. We
have considered this term, as well as a weak  $\tilde{J}_{pd}$
coupling between the doped hole occupying a non-ligand orbital and its
neighbor Cu spins (FM exchange is favored by Hund's coupling when the
Cu hole hops into the O orbital orthogonal to that hosting the doped
hole). None of these terms were found to lead to qualitative
changes, and because of their relatively small magnitude, their
quantitative effects are minor.
\end{thebibliography}
\end{document}